\documentclass{iau} 
\usepackage{graphicx}

\newcommand{\apj}{{\it ApJ}}
\newcommand{\solphys}{{\it Sol.~Phys.}}

\newcommand{\apjl}{{\it ApJ Letters}}

\title[JD 11.~~Data Handling] %% give here short title %%
{Data Handling and Assimilation for\\ Solar Event Prediction}

\author[Petrus C. Martens \& Rafal A. Angryk]   %% give here short author list %%
{Petrus C. Martens$^1$ \and Rafal A. Angryk$^2$}

\affiliation{$^1$Department of Physics \& Astronomy,
Georgia State University,\\
25 Park Place, 6$^{th}$ Floor,
Atlanta, GA 30303, USA\\ 
email: {\tt martens@astro.gsu.edu}\\[\affilskip]
$^2$Department of Computer Science,
Georgia State University,\\
25 Park Place, 7$^{th}$ Floor,
Atlanta, GA 30303, USA\\ 
email: {\tt angryk@cs.gsu.edu}}

\pubyear{2018}
\volume{335}  %% insert here IAU Symposium No.
\jname{Space Weather of the Heliosphere: Processes and Forecasts}
\editors{C. Foullon \& O. Malandraki, eds.}
\begin{document}

\maketitle

\begin{abstract}

The prediction of solar flares, eruptions, and high energy particle storms is of
great societal importance.  The data mining approach to forecasting has been
shown to be very promising.  Benchmark datasets are a key element in the
further development of data-driven forecasting.  With one or more benchmark
data sets established, judicious use of both the data themselves and the 
selection of prediction algorithms is key to developing  a high quality and
robust method for the prediction of geo-effective solar activity.  We review here
briefly the process of generating benchmark datasets and developing prediction
algorithms.

\end{abstract}

\firstsection % if your document starts with a section,
\section{Introduction:  Flare Prediction With Machine Learning}

Forecasting the timing and magnitude of geo-effective flares, CMEs, and
Solar Energetic Particle (SEP) events is of great practical significance for
our increasingly technologically dependent society, and hence research in
the field of space weather is strongly supported in many developed nations.
For example the US has developed a National Space Weather Strategy
and Space Weather Action Plan\footnote{https://goo.gl/r0Vcd5} 
that gives priority to research on the forecasting of solar
flares, eruptions, and particle storms, while also emphasizing the need to
develop benchmarks, the subject of this paper.

One approach to flare and SEP forecasting is using the methods of Big
Data Mining, also called Machine Learning.  The method involves gathering
large amounts of relevant data, then cleaning up and
balancing these data, as described in detail below, to produce a ``benchmark 
data set" that algorithms of data analytics can be safely applied on without
generating spurious results.  That procedure, developed and widely used in
Computer Science, is still only sparsely used in Solar Physics.  The purpose
of this paper is to specify and explain this procedure for the solar physics
community.

The record for this method in the prediction of flares and SEP events 
gives us
reason to be optimistic.  First of all, experienced human flare forecasters 
at NOAA's Space Weather Prediction Center have used a similar method
for decades with good outcomes.  Secondly,
recent research by Falconer {\em et al.} (2014), Bobra \& Couvidat (2015),
as well as our own work (Angryk {\em et al.}, 2018, in preparation),
using data from SDO's Helioseismic and Magnetic Imager (HMI; Scherrer
{\em et al.}, 2012), has yielded strong results for flare forecasting.
\nocite{falconer-etal14, bobra-couvidat15, scherrer-etal12}

\section{Development of a Benchmark Data Set}

The data repositories for most solar observatories are what we call
``Operational Archives",
meaning that all the data collected are archived, to be sorted out for
usefulness by researchers as they analyze them.  For most observatories
the images undergo substantial processing -- e.g. background subtraction,
North up, etc. -- prior to release for analysis.  For the data from NASA's
Solar Dynamics Observatory (SDO), also extensive metadata are derived,
mostly from automated modules developed by the SDO Feature Finding
Team (FFT; Martens {\em et al.}\ 2012) and stored in the Heliophysics Events
Knowledge base (HEK; Hurlburt {\em et al.}\ 2012).  These metadata too
usually become part of a benchmark data set; in fact the analysis from most
benchmarks is performed on metadata only.
\nocite{martens-etal12, hurlburt-etal12}

Since the benchmark is being developed for {\em automated}
analysis, it is important first to {\em clean} the data, i.e.\  remove all the
images that are compromised in any way, for example by CCD bleeds,
proton storm hits on the CCD, or from electronic garbling.  Operational
archives often have been labeled for such cases.
The second step is to register all data gaps.  Fig.~1 (Schuh {\em et al.},
2015) shows the identification of data gaps in the metadata derived by
our trainable Content Based Image Recognition (CBIR) module (Martens
{\em et al.}, 2012), using all channels of the Atmospheric Imaging
Assembly (AIA; Lemen {\em et al.}, 2012) on SDO.  The data gaps are
either in the AIA data or from the module itself not operating properly.
Note the bi-annual gaps when SDO is eclipsed by Earth.
\nocite{lemen-etal12}

\begin{figure}[ht]
% \vspace*{-2.0 cm}
\begin{center}
\includegraphics[width=1.00\linewidth]{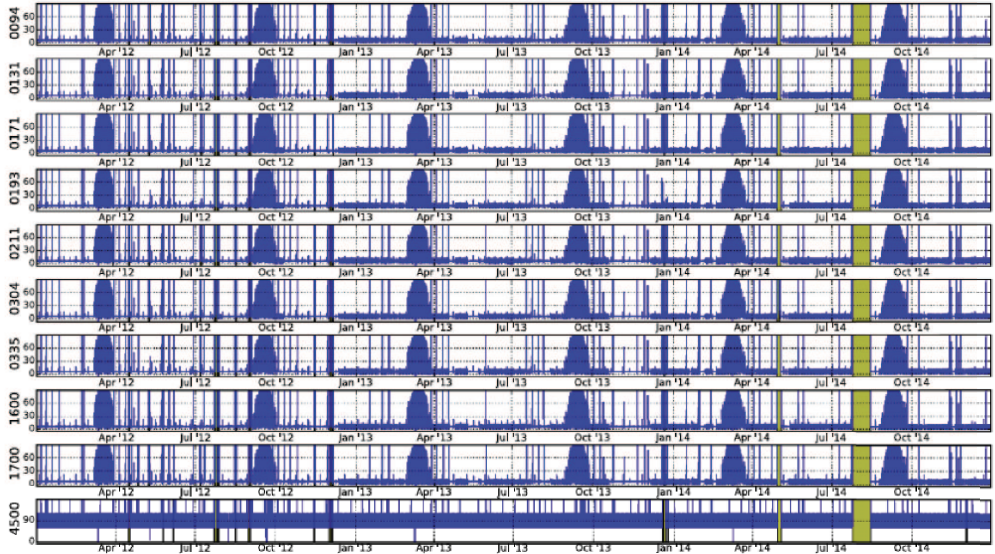}
% \vspace*{-1.0 cm}
 \caption{Time difference in minutes between metadata entries produced
by our trainable CBIR module 
for each AIA channel for three years of AIA data.  Yellow shows a
gap of over 12 h.  Figure from Schuh, Angryk \& Martens (2015).
   \label{fig1}}
\end{center}
\end{figure}

Why is it necessary to catalog these gaps for the benchmark data set?
We'll give one example.  Say, one does a study of the occurrence of
sigmoids for the prediction of flares.  For automated sigmoid detection we
have the so-called ``sigmoid-sniffer" (Martens {\em et al.}, 2012), developed
by Dr.\ Georgoulis, and for flares we use the X-ray data from the GOES 
satellites\footnote{https://satdat.ngdc.noaa.gov/sem/goes/data/new\_avg/}.
For the study
results to be reliable one must exclude the periods when there are no AIA,
or only low cadence
94 \AA\ data that the ``sigmoid sniffer" uses, as well as any intervals for which
no GOES flare data are available in the pertinent X-ray channel, and also
any timespans for which the ``sigmoid sniffer" itself has no recorded
metadata.  Without that knowledge any classifier algorithm would be misled
and produce incorrect results.
Solar physicist conducting their own studies do this data cleaning by
themselves and by hand, which is one reason why such studies often are
limited to several dozen events -- more would require a far too large effort.

The purpose then of our group in developing benchmark datasets is twofold. 
First by reviewing large amounts of data, as exemplified for AIA in Fig.~1, we
can develop clean, properly labeled datasets with thousands of data points.
This entails an enormous amount of work -- more than a year by a
graduate student in the Schuh, Angryk \& Martens (2015 and 2016) and
Kucuk, Banda, \& Angryk (2017) papers.   The latter released a dataset
of automatically generated records of ARs, Coronal Holes, Sigmoids,
and Flares observed by AIA, together with their image data, with 270,000
event records. It would be truly a wasteful multiplication of effort to repeat
such a program over and over again for different research projects, hence
our benchmark  datasets.
\nocite{schuh-etal15, schuh-etal16, Kucuk2017}

Secondly, and at least as important, our
benchmark can then be used to make fair comparisons between different
methods to answer the same research question.  Consider the prediction of
solar flares:  the ``skill" scores that
any prediction algorithm achieves will obviously depend on the dataset that
one uses to train and test the prediction algorithm
on.  Hence comparisons between different algorithms honed on different
data sets are rather meaningless; a fair comparison is achieved only by
comparing the ``skill scores" of different algorithms when applied to the
same dataset.  This is one of the reasons why the US Space Weather Action
Plan, cited above, specifically asks for benchmarks to
be developed.  Obviously benchmarks must be ``Static Archives",
that is, once established they should not be changed to make the
analyses carried out with them comparable.

Since the launch of SDO up to July 2017 we found 580 detected X- and
M-class flares, leaving us with a still fairly small dataset to train and
test predictions on, but still enough to obtain promising results as Bobra \& Couvidat
(2015) have shown, and as our own work indicates (Angryk {\em et al.}, in
preparation).  The number of geo-effective SEP events is less then one hundred,
and hence we use a somewhat different approach here, that of {\em instance
based learning}; see Filali {\em et al.}\ (2017).

\section{The Need for Balanced Training Sets}  

When developing a flare prediction algorithm one needs to consider not only
recorded flare events but also a similar amount of {\em non-events}, that is, 
instances of ARs that did {\em not} flare in the forecast window.  Solar ARs
do not have X- and M-class flares most of the time. There have been $\approx$
600 in the six years of the SDO mission, that is roughly one every
four days over the entire disk, and a considerably longer cadence per individual
AR.  Hence there are far more non-flaring daily intervals for each AR than there
are flaring instances.  Considering all these instances together will lead to a
very uneven distribution.  Any sophisticated flare prediction algorithm may then
simply fall back to an ``All Clear" prediction all of the time for each AR, to get
the correct outcome in the great majority of cases.  This is meaningless of
course since the whole point of the exercise is to predict these rare flares.

One standard solution is to reduce the number of non-flaring intervals in the
training data derived from the benchmark dataset by a large factor, for 
example by picking a random subsample.  It is important to be very careful
here to not introduce a systematic bias.  A truly random sampling of non-flaring
intervals may be the safest solution here, but it leaves us with a rather small
dataset for training and testing, only about 1200 events, 600
flaring and 600 non-flaring intervals.

An alternative method to address imbalance is to identify forecasting evaluation
measures that are resistant to imbalanced data.  One can use the Heidke
Skill Score (HSS) (e.g.\ Barnes \& Leka, 2008) or the Gilbert Skill score (GS;
e.g.\  Mason \& Hoeksema, 2010). 
\nocite{barnes-leka08, mason-hoeksema10}

\section{Using Classifiers}

The next step is applying machine learning methods to the benchmark dataset,
in our case to find the best possible flare prediction. 
Almost all data mining papers on flare forecasting thus far fall into the category of 
parameter-based classification, which transforms spatio-temporal solar image data
into a time series representation.  Usually a group of  expert-picked parameters is
extracted from the images, that summarize the evolution of these images.  For
example, Bobra \&  Couvidat
(2015) use 25 physical parameters derived from vector-magnetograms.  Thus
the project is reduced to the study of time series of metadata, sometimes already
published online by the instrument teams.  

One then applies conventional classification methods, such as regression models,
decision trees, support vector machines, neural networks, k-nearest neighbor, etc.\
on the extracted parameters. The classifier looks back over a given time-interval
to come up with a prediction for a second time interval, the prediction window.
Once these classifiers are trained on one
part of the benchmark dataset (usually of the order of 2/3 of the data), they are
tested for their forecasting capability on the remaining part of the benchmark data.
After that one picks the best, or a combination of the best algorithms, to employ them
in an operational setting to predict real-time solar activity.  That is the second
and decisive test of the capabilities of the system.

A key aspect of this procedure is that one has to have a strong understanding
of how the classifiers that one applies work, to figure out when true machine
learning, i.e.\ data generalization, occurs. {\em When you encounter a black box -- 
look inside it!}  Otherwise fundamental mistakes will be made.  Even a basic 
undergraduate textbook like {\em Principles of Data Mining} (Bramer 2013, second
edition) can be very helpful.  Another, even more effective approach is the close
collaboration between solar physicists and computer scientists, as we have in the
recently established
interdisciplinary Solar-stellar Informatics Cluster at Georgia State University.
\nocite{bramer13}

%% Get references from Bib-file
%\renewcommand\refname{}
\bibliographystyle{apj}
%\bibliography{references}

\end{document}